\documentclass[ aip, rsi, amsmath,amssymb, reprint, ]{revtex4-1}
\usepackage{graphicx}
\usepackage{dcolumn}
\usepackage{bm}
\usepackage{amssymb}
\usepackage{amsmath}
\usepackage{amsfonts}

\usepackage{hyperref}
\usepackage[usenames]{color}

\begin{document}


\title{New type of chimera and mutual synchronization of spatiotemporal structures in two coupled 
ensembles of nonlocally interacting chaotic maps}

\author{Andrei Bukh}
\email{buh.andrey@yandex.ru}
\affiliation{Department of Physics, Saratov State University, Astrakhanskaya str. 83, 410012 Saratov, 
Russia}
\author{Elena Rybalova}
\email{rybalovaev@gmail.com}
\affiliation{Department of Physics, Saratov State University, Astrakhanskaya str. 83, 410012 Saratov, 
Russia}
\author{Nadezhda Semenova}
\email{semenovani@info.sgu.ru}
\affiliation{Department of Physics, Saratov State University, Astrakhanskaya str. 83, 410012 Saratov, 
Russia}
\author{Galina Strelkova}
\email{strelkovagi@info.sgu.ru}
\affiliation{Department of Physics, Saratov State University, Astrakhanskaya str. 83, 410012 Saratov, 
Russia}
\author{Vadim Anishchenko}
\email{wadim@info.sgu.ru}
\affiliation{Department of Physics, Saratov State University, Astrakhanskaya str. 83, 410012 Saratov, 
Russia}

\date{\today}

\begin{abstract}
We study numerically  the dynamics of a network made of two coupled one-dimensional ensembles of 
discrete-time systems. The first ensemble is represented by a ring of nonlocally coupled Henon maps, and 
the second one - by a ring of nonlocally coupled Lozi maps. We find that the network of coupled 
ensembles can realize all the spatio-temporal structures which are observed both in the Henon map 
ensemble and in the Lozi map ensemble when uncoupled. Moreover, we reveal a new type of spatiotemporal 
structure, a solitary state chimera, in the considered network. We also establish and describe the 
effect of mutual synchronization of various complex spatiotemporal patterns in the system of two coupled 
ensembles of Henon and Lozi maps.

\end{abstract}

\maketitle

\begin{quotation}
Recently studying the formation and evolution of various spatiotemporal patterns in ensembles or 
networks of coupled oscillators has become one of the most rapidly developing and highly attractive 
research topics in the nonlinear science  and its applications. This exclusive interest is especially 
related to the discovery of a novel type of spatiotemporal structure -- a chimera state. A lot of 
attention is paid to the dynamics of coupled ensembles of identical elements with various
coupling topologies, but of particular interest are coupled ensembles with different 
types of network elements. In the latter case, the enrichment of regimes as well as the 
synchronization of spatiotemporal patterns is expected to be observed. In the present paper we 
analyze the spatiotemporal dynamics of a network made of two coupled rings of Henon and Lozi 
maps with 
nonlocal coupling. Our numerical studies have shown that this network can demonstrate both the 
spatiotemporal
regimes, which are observed in separate rings, and a new type of chimera structure, called a solitary 
state chimera. We have also established the possibility of realizing the mutual synchronization of 
various complex spatiotemporal structures in the network of two coupled rings. The identity of 
synchronous patterns is confirmed by calculating the cross-correlation coefficient. The existence of a 
finite region of synchronization in the coupling 
parameters plane of the considered system is shown for an exemplary synchronous structure. 
\end{quotation}

\section{Introduction}
\label{part_intro}

Studying the dynamics of complex ensembles of coupled nonlinear oscillators has been the subject of 
intensive research  for many years 
\cite{afraimovich1995stability,osipov2007synchro,belykh2001cluster,nekorkin2011,nekorkin2002synergetic}.
At present, a special attention in this research direction is targeted to the analysis of 
formation and evolution of a novel type of spatiotemporal patterns, a so-called chimera state 
\cite{Kuramoto-2002,Abrams-2004}.
The chimera state represents the coexistence of  clearly identified clusters of 
oscillators with asynchronous (incoherent) and synchronous (coherent) dynamics. 
The efforts of many researchers are aimed at searching for  new types of chimera structures,   
at describing their dynamical and statistical characteristics, and  at the ability to give their 
justified classification 
\cite{Omelchenko-2011,Wolfrum2011,Zakharova-2014,Semenova2016PRL,Scholl2016EPJST,Anishchenko-CNSNS2016,
ShepelevCNSNS2018,Kemeth-2016}.
Up to now, several types of chimeras have been revealed, for example, phase and amplitude chimeras, 
coherence resonance chimera, double-well chimera, nonstationary chimeras of switching type, and others  
\cite{Omelchenko-2011,Semenova2016PRL,Anishchenko-CNSNS2016,ShepelevCNSNS2018,Semenova2017Temporal}.
Chimera structures have been found in ensembles of coupled oscillators of a very different nature with 
regular and chaotic dynamics \cite{Semenova2016PRL,Scholl2016EPJST,Shepelev2017PLA,Shepelev2017ND}.
Besides, chimeras can be realized and observed both numerically and experimentally 
\cite{Hagerstrom-NaturePhys-2012,Tinsley-NaturePhys-2012,Larger-PRL-2013,Schmidt-Chaos-2014}.
 It is important to note that the majority of works is devoted to  the study of one- or 
two-dimensional networks of identical oscillators with different coupling topology between them. 
It appears to be very interesting to explore the dynamics of coupled ensembles in the case when 
the individual ensembles consist of elements of  different types  
\cite{Boccaletti-2014-PhysRep,Semenova2015EPL,Rybalova2017EPJST}. In such more complicated 
networks  one can expect the complex cooperative dynamics and the appearance of new 
interesting incoherence (chimera) states \cite{ALSG2015,ALSG2016,Perc2017}. Moreover, one can naturally 
formulate and study the problem of synchronization of various spatiotemporal structures  in networks of 
coupled ensembles \cite{osipov2007synchro,Boccaletti-2014-PhysRep,Hovel}. 

In the present paper we explore the spatiotemporal dynamics of a network which consists of two coupled 
one-dimensional ensembles of discrete-time systems. 
The first ensemble is represented by a ring of nonlocally coupled chaotic Henon maps, and the second 
one -- by a ring of nonlocally coupled chaotic Lozi maps. It is important to mention that the first 
network elements 
possess nonhyperbolic chaotic attractors, while the second network elements are characterized by 
hyperbolic chaotic 
attractors. In the papers \cite{Semenova2017Temporal,Semenova2015EPL,Rybalova2017EPJST} it has 
been shown that the one-dimensional ensemble of nonlocally coupled Henon maps can realize phase, 
amplitude 
and nonstationary (temporally intermittent) chimeras, while the ensemble of nonlocally coupled Lozi maps 
can demonstrate only solitary state structures and no chimeras appear. In this connection we are 
especially  
interested in studying various types of spatiotemporal structures and their possible mutual 
synchronization in a network of two coupled one-dimensional networks of the types indicated above.

\section{Mathematical Model of the System under Study}
\label{part_math}

\begin{figure}
\begin{center}
\includegraphics[width=.9\columnwidth]{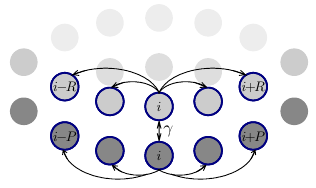}
\end{center}
\caption{Schematic representation of the considered network and its coupling topology.}
\label{two_coupled_rings}                                                                                                   
\end{figure}

In our research we explore the network of two interacting 
one-dimensional rings of Henon and Lozi maps. When uncoupled, each of the ensembles is  
characterized by nonlocal 
coupling between their individual elements. These two rings are coupled in the way as shown in 
Fig.~\ref{two_coupled_rings}.
Each $i$th element of the first ensemble is coupled with the $i$th element of the second one with 
the coupling coefficient $\gamma$. The network  of two coupled ensembles   is described by the following 
equations:
\begin{equation}\label{main_eq}
\begin{aligned}
   & x_i^{t+1} = f_i^t + \dfrac{\sigma_1}{2P} \sum_{j=i-P}^{i+P} \left[ f_j^t - f_i^t\right] + \gamma F_i^t,
\\ & y_i^{t+1} = \beta x_i^t,
\\ & u_i^{t+1} = g_i^t + \dfrac{\sigma_2}{2R} \sum_{j=i-R}^{i+R} \left[ g_j^t - g_i^t\right] + \gamma G_i^t,
\\ & v_i^{t+1} = \beta u_i^t.
\end{aligned}
\end{equation}
Here  $i=1,2,\dots,N$ is the element number in each ring,  $N=1000$ is the total number of elements in 
each ensemble, $\sigma_1$ and $\sigma_2$ are the nonlocal coupling strengths in the Henon and Lozi map 
rings, respectively.  $P$ and $R$ denote the number of couplings of the $i$th element with its nearest 
neighbors from each side in the Henon and Lozi map ensembles, respectively, and  $\gamma$ is the 
coupling coefficient between  two elements from different ensembles. 

The first pair of equations in the system \eqref{main_eq} describes the ring of nonlocally coupled Henon 
maps, where $f_i^t=f(x_i^t,y_i^t)=1-\alpha(x_i^t)^2+y_i^t$. The number of neighbors $P$ 
and the coupling strength  $\sigma_1$ are set in such a way that, without coupling between the two 
rings, the Henon map ensemble demonstrates a chimera state. 

The second pair of equations in \eqref{main_eq} determines the ring of nonlocally coupled Lozi maps, 
where  $g_i^t=g(u_i^t,v_i^t)=1-\alpha|u_i^t|+v_i^t$. The number of neighbors $R$ and the 
coupling strength $\sigma_2$ are chosen so that the regime of solitary states is realized in the Lozi map 
ensemble when the two rings are uncoupled. The control parameters of  both networks' elements are fixed 
as $\alpha=1.4$ and $\beta=0.3$, that corresponds to the chaotic behavior in all the individual elements 
of the considered system. 

Functions $F_i^t$ and $G_i^t$ in the network~\eqref{main_eq} are responsible for the coupling type
between the two ensembles. In our numerical simulation  we deal with  dissipative and inertial coupling 
types. We have for the  dissipative coupling 
\begin{equation}
\label{dissipative_eq}
\begin{aligned}
   & F_i^t = {\bar F}_i^t (x_i^t,y_i^t,u_i^t,v_i^t)=g(u_i^t,v_i^t)-f(x_i^t,y_i^t),
\\ & G_i^t = {\bar G}_i^t (x_i^t,y_i^t,u_i^t,v_i^t)=f(x_i^t,y_i^t)-g(u_i^t,v_i^t),
\end{aligned}
\end{equation}
and for the inertial coupling
\begin{equation}
\label{inerz_eq}
\begin{aligned}
   & F_i^t = \bar {\bar F}_i^t (x_i^t,y_i^t,u_i^t,v_i^t)=u_i^t-x_i^t,
\\ & G_i^t = \bar {\bar G}_i^t (x_i^t,y_i^t,u_i^t,v_i^t)=x_i^t-u_i^t.
\end{aligned}
\end{equation}
These coupling type definitions correspond to the interaction character of coupled rings. 
 The dissipative coupling tends to make equal the instantaneous amplitudes of  coupled elements, 
 while the inertial coupling possesses the ability to  keep (retain) the memory about the previous state 
of the system.

\section{Numerical results}
\label{part_numeric_simulation}

\begin{figure}
\begin{tabular}{cc}
\includegraphics[width=.45\columnwidth]{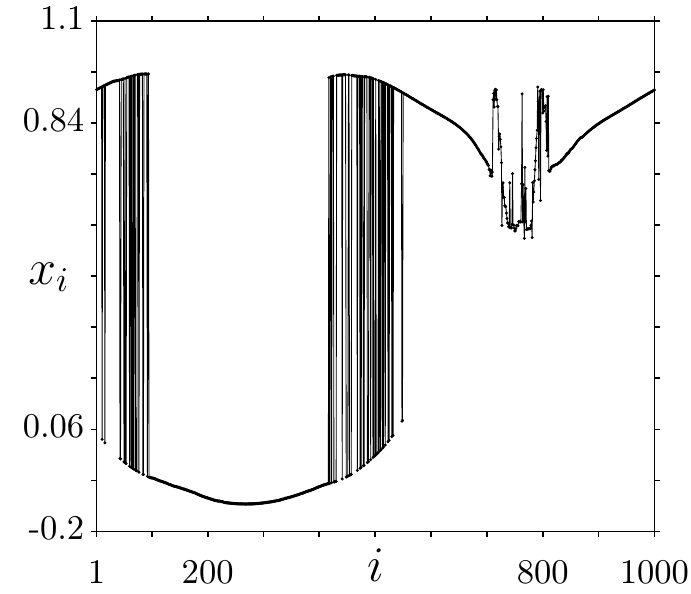} &
\includegraphics[width=.45\columnwidth]{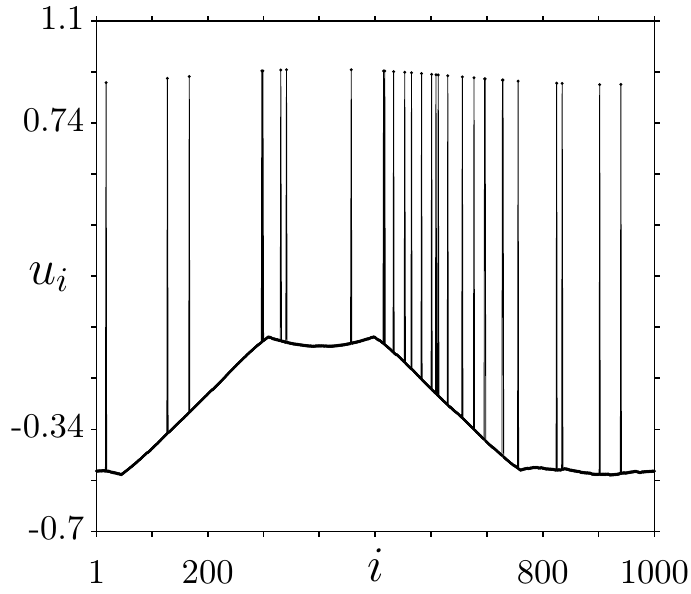} \\
\hspace{8pt} (a) & \hspace{8pt} (b)
\end{tabular}
\caption{Instantaneous profiles of amplitudes (snapshots) in the network \eqref{main_eq} without coupling
between  the rings ($\gamma=0$) for  the Henon map ensemble at  $\sigma_1=0.32$ and $P=320$ (a), and 
the Lozi map ensemble at $\sigma_2=0.225$ and $R=190$ (b). }
\label{smooth_profs_gamma_0}
\end{figure}

As can be seen from ~\eqref{main_eq}, we have two uncoupled ensembles of Henon and Lozi maps when  
$\gamma=0$. 
In this case, the Henon map network can demonstrate two types of chimera structures -- phase and 
amplitude chimeras. They can be observed, for example, with $\sigma_1=0.32$ and 
$P=320$, as was shown in the papers \cite{Rybalova2017EPJST,SemenovaRCD2017}. These 
structures are exemplified in 
Fig.~\ref{smooth_profs_gamma_0}(a). The phase chimera represents the coexistence of a coherence cluster 
 ($100 \leqslant i \leqslant 400$) and  incoherence clusters of elements ($1 \leqslant i \leqslant 100$ 
and $400 \leqslant i \leqslant 550$) with periodic dynamics and irregular phase shifts relative to each 
other. Consequently, the cross-correlation coefficient $R_{1,i}$ for the phase chimera is equal to unity 
in modulus but changes irregularly its sign ($R_{1,i}=\pm1$) inside the incoherence cluster. 
The spatial cross-correlation coefficient   $R_{1,i}$ is estimated for the elements of the 
spatiotemporal structures realized in one of the considered ensembles, as was described in the 
paper \cite{Chaos2016correlation}:
\begin{equation}
 R_{1,i}=\dfrac {\langle \tilde x_1 (t) \tilde x_i (t) \rangle}
{\sqrt{\langle \tilde x_1^2 (t) \rangle 
\langle \tilde x_i^2  (t) \rangle}}, \quad i=2,\ldots, N,
\label{CCC}
\end{equation}
where $\tilde{x}=x(t)-\langle x(t)\rangle$. The amplitudes of oscillations  $x(t)$ in \eqref{CCC} 
correspond to the values of  the $x$ (or $u$) coordinate of one of the ensembles.

The elements belonging to the amplitude chimera ($700 \leqslant i \leqslant 820$ in 
Fig.~\ref{smooth_profs_gamma_0}(a))
oscillate strongly chaotically and are uncorrelated to each other. Besides, this chaotic regime is 
characterized by temporal intermittency and the lifetime of the amplitude chimera is finite 
\cite{Semenova2017Temporal}. 

As was described in the papers \cite{Rybalova2017EPJST,SemenovaRCD2017}, the ring of nonlocally coupled 
Lozi maps demonstrates the regime of 
solitary states. This spatiotemporal structure is displayed in Fig.~\ref{smooth_profs_gamma_0}(b) for  
$\sigma_2=0.225$ and $R=190$. 

\subsection{Spatiotemporal structures in the network of coupled ensembles}

\begin{figure}
\begin{tabular}{cc}
\includegraphics[width=.45\columnwidth]{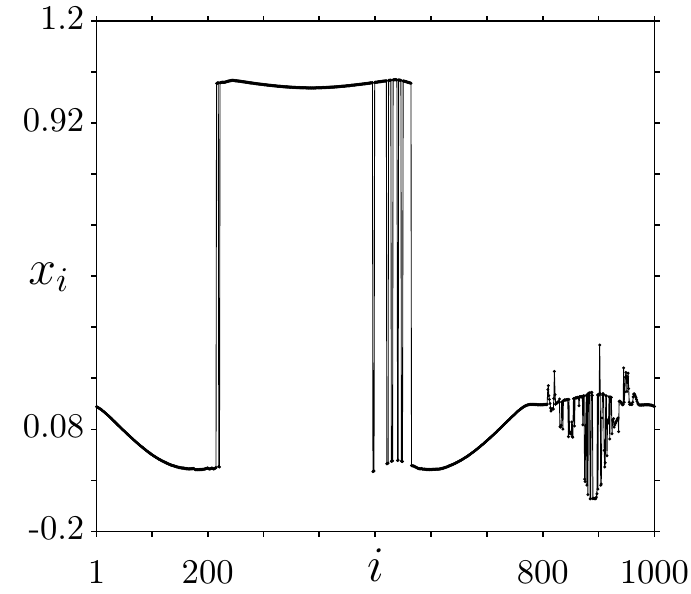} &
\includegraphics[width=.45\columnwidth]{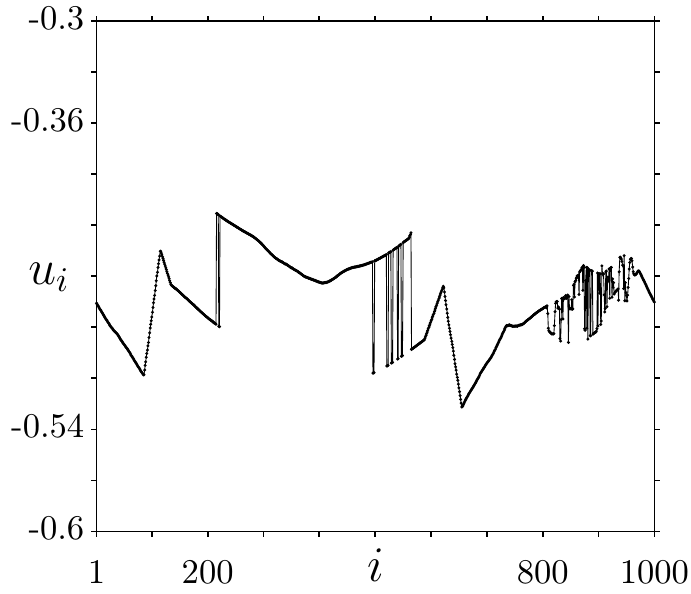} \\
\hspace{8pt} (a) & \hspace{8pt} (b)
\end{tabular}
\caption{Snapshots for the network~\eqref{main_eq} dynamics in the case of dissipative coupling 
\eqref{dissipative_eq}
for (a) the Henon map ensemble and (b) the Lozi map ensemble. Parameters: $\sigma_1=0.32$, 
$\sigma_2=0.225$, $P=320$, $R=190$, and $\gamma=0.015$. 
The amplitude chimera includes the elements $800\leqslant i\leqslant 980$, and the phase chimera 
consists of two clusters: $210\leqslant i\leqslant230$ and $490\leqslant i\leqslant570$  in both rings.}
\label{smooth_profs_diss_gamma_0_115}
\end{figure}

We now consider the dynamics of the network of two coupled ensembles~\eqref{main_eq} with the 
dissipative coupling
\eqref{dissipative_eq} for finite values of $\gamma>0$. Instantaneous profiles (snapshots) for the states 
in both rings 
are shown in Fig.~\ref{smooth_profs_diss_gamma_0_115} for $\gamma=0.115$. Our numerical calculations have 
indicated that if the two rings are dissipatively coupled, each of them can demonstrate phase and 
amplitude chimeras and solitary states. With this, for instance, the Lozi map ensemble can exhibit the 
regime which cannot be observed without coupling with the Henon map ring. As can be seen from  
Fig.~\ref{smooth_profs_diss_gamma_0_115}(b), the amplitude ($800 \leqslant i \leqslant 980$ ) and  
phase ($210 \leqslant i \leqslant 230$) and ($490 \leqslant i \leqslant 570$) chimeras are realized in the Lozi map ring.

\begin{figure}
\begin{tabular}{cc}
\includegraphics[width=.45\columnwidth]{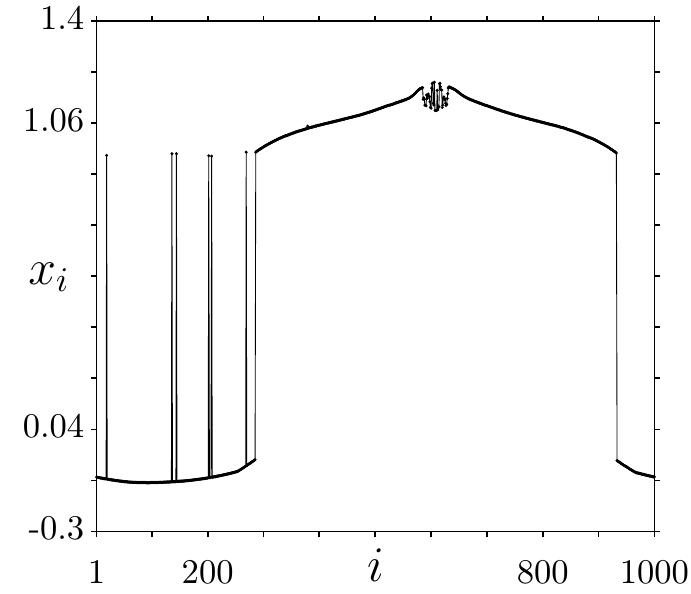} &
\includegraphics[width=.45\columnwidth]{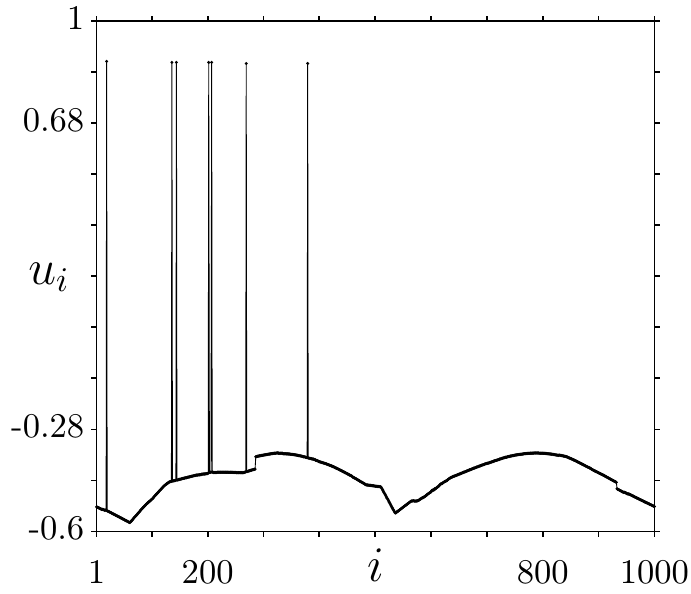} \\
\hspace{8pt} (a) & \hspace{8pt} (b)
\end{tabular}
\caption{Solitary state structure in the Henon map ensemble of the network~\eqref{main_eq} for the  case 
of dissipative coupling \eqref{dissipative_eq}. (a) Snapshot of the dynamics of the Henon map ensemble, 
and (b) snapshot of the dynamics of the Lozi map ensemble. Parameters: $\sigma_1=0.35$, 
$\sigma_2=0.225$, $P=320$, $R=190$, and $\gamma=0.0085$.}
\label{smooth_profs_diss_gamma_0_xxx}
\end{figure}

In a similar manner, solitary state structures can be realized in the Henon map ensemble, while they 
cannot
be observed in this ensemble without coupling with the Lozi map ring. The regime of solitary states in 
the Henon map ensemble is exemplified in Fig.~\ref{smooth_profs_diss_gamma_0_xxx}(a) for the dissipative
coupling \eqref{dissipative_eq} with the Lozi map ensemble. As follows from this figure, the solitary 
states are realized within the interval of elements $1 \leqslant i \leqslant 280$, and the amplitude 
chimera occupies the elements  $570 \leqslant i \leqslant 640$. The corresponding snapshot of the 
dynamics of the Lozi map ensemble is shown in Fig.~\ref{smooth_profs_diss_gamma_0_xxx}(b).

It has been shown in the paper \cite{Semenova2017Temporal}  that the lifetime of the amplitude chimera 
in 
 the ensemble of nonlocally coupled Henon map is finite and can be indefinitely increased by 
the influence of noise. If the amplitude chimera regime is established in the system \eqref{main_eq} of 
coupled ensembles of Henon and Lozi maps (Fig.~\ref{smooth_profs_diss_gamma_0_115}), then its lifetime 
can be controlled by varying the coupling coefficient $\gamma$ between the ensembles. Our calculations 
have shown that the dependence of the lifetime on $\gamma$ is essentially nonlinear and may vary 
over a wide range, including regimes when the lifetime grows  infinitely. This effect is achieved in the 
autonomous system \eqref{main_eq} without external influences. 

In the case of inertial coupling \eqref{inerz_eq} in \eqref{main_eq} and for certain values of the 
coupling parameters,
the coupled ensembles of Henon and Lozi maps can also demonstrate new spatiotemporal patterns which 
cannot be realized when they are uncoupled  ($\gamma=0$). 
For example, the Henon map ring  can exhibit the regime of traveling waves, which is not observed 
without coupling with the Lozi map ensemble.

\subsection{Solitary state chimera}

\begin{figure}
\begin{tabular}{cc}
\includegraphics[width=.45\columnwidth]{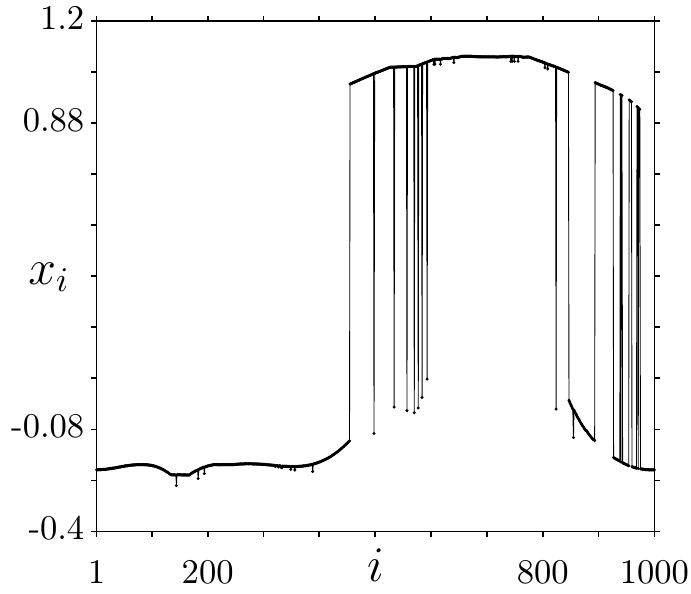} &
\includegraphics[width=.45\columnwidth]{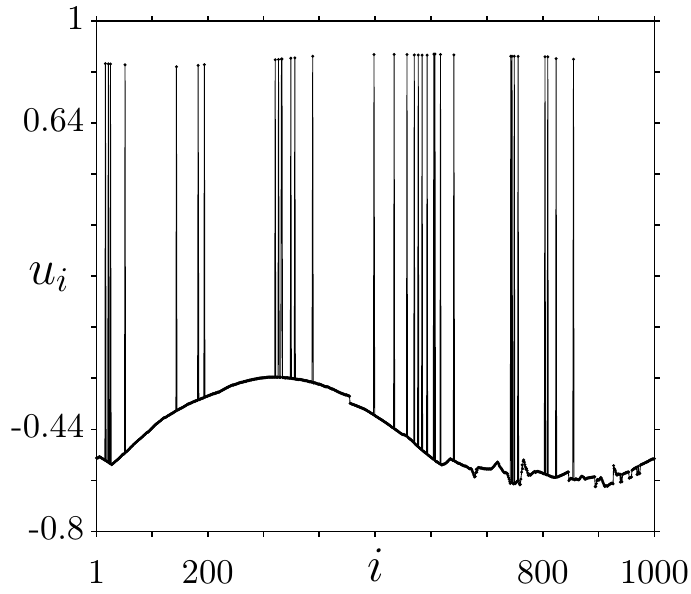} \\
\hspace{8pt} (a) & \hspace{8pt} (b) \\
\includegraphics[width=.45\columnwidth]{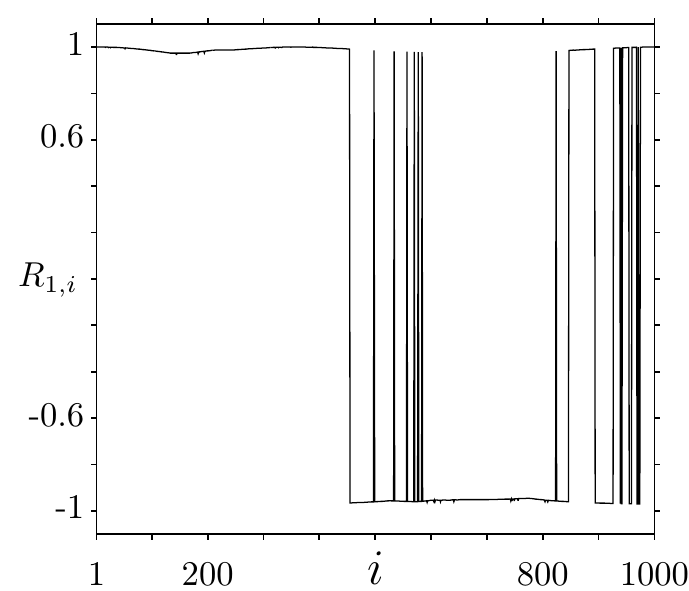} &
\includegraphics[width=.45\columnwidth]{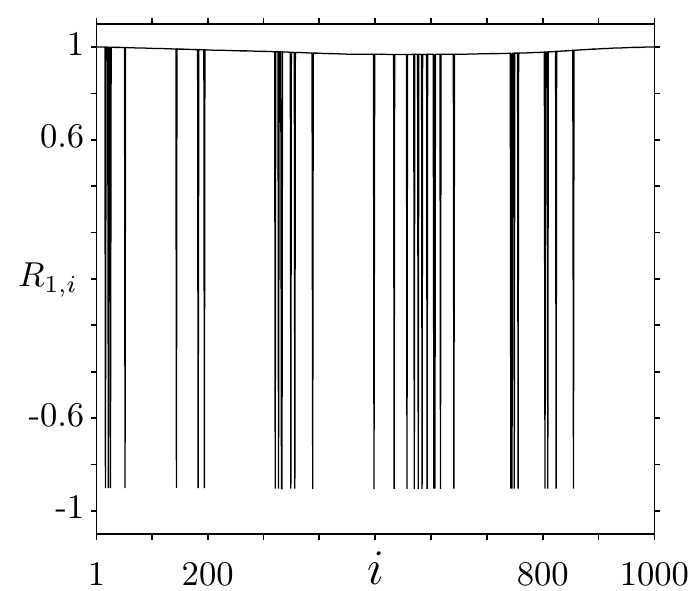} \\
\hspace{8pt} (c) & \hspace{8pt} (d)
\end{tabular}
\caption{Snapshots of the $x_i$ and $u_i$ states and cross-correlation coefficient $R_{1,i}$
in the network \eqref{main_eq} for the inertial coupling \eqref{inerz_eq} for  (a,c) solitary state 
chimera in the Henon map ensemble, and (b,d) solitary states in the Lozi map ensemble. The solitary state 
chimera forms the cluster with elements $490\leq i \leq 590$ in Fig.~\ref{solitary_state_chimera}(a). 
Parameters: $\sigma_1=0.34$, $\sigma_2=0.205$, $P=320$, $R=193$, and $\gamma=0.02$.}
\label{solitary_state_chimera}
\end{figure}

Now we would like to pay attention to one more 
spatiotemporal structure which is found in the coupled rings and can be referred to as a {\it solitary 
state chimera}. This new structure is observed in the Henon map ring 
(Fig.~\ref{solitary_state_chimera}(a)), while the solitary state regime is detected in the Lozi map 
ensemble (Fig.~\ref{solitary_state_chimera}(b)). The newly found spatiotemporal mode can be classified 
as a chimera state, since this incoherence cluster, which includes the network elements with a different 
behavior (as compared with the other elements),  is strongly localized in the space. 
The main differences between the new structure and the phase chimera consist in the facts that the 
elements from the solitary state chimera represent solitary states and demonstrate 
asynchronous chaotic oscillations. Moreover, the coherence domain corresponds to complete chaotic 
synchronization, while we deal with periodic or weakly chaotic behavior in the case of phase chimera. 

Our finding can be confirmed quantitatively by estimating the cross-correlation coefficient 
\eqref{CCC} for the new chimera structure. 
The calculation results are presented in Fig.~\ref{solitary_state_chimera}(c,d) for the Henon map 
ensemble and the Lozi map ensemble, respectively. As can be seen from 
Fig.~\ref{solitary_state_chimera}(c), all the values of $R_{1,i}$ do not achieve unity in absolute value 
($|R_{1,i}|<1$) and this corroborates the fact that all the elements in the solitary state chimera 
behave chaotically. 
In the case of phase chimera the values of $R_{1,i}$ are irregularly switched (alternated) between 
two levels $\pm 1$.

\begin{figure}
\begin{tabular}{cc}
\includegraphics[width=.45\columnwidth]{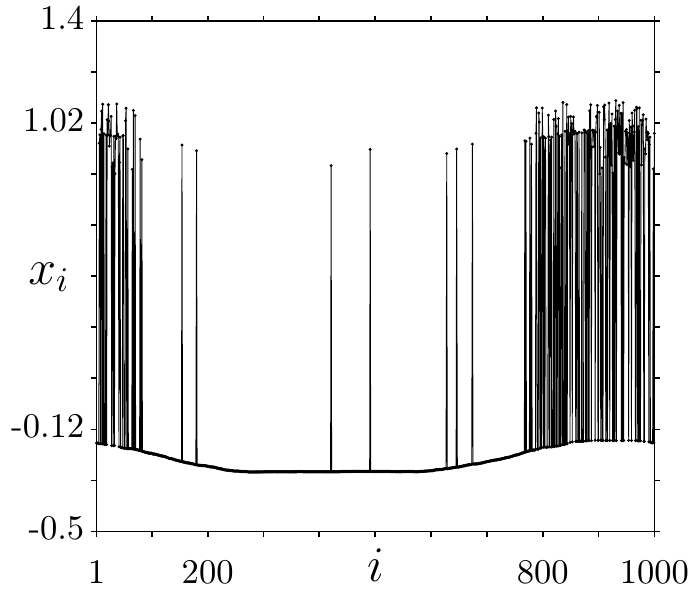} &
\includegraphics[width=.45\columnwidth]{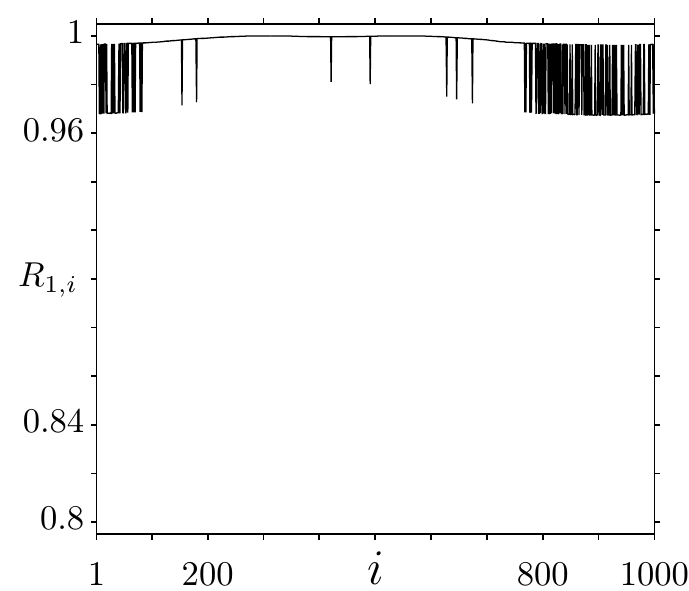} \\
\hspace{8pt} (a) & \hspace{8pt} (b)
\end{tabular}
\caption{Snapshot of the variables $x_i$ (a) and the cross-correlation coefficient  $R_{1,i}$ 
(b) for the dissipative coupling \eqref{dissipative_eq} in the network \eqref{main_eq} for the solitary 
state chimera in the ensemble of Henon maps. This chimera consists of the elements  $1<i<100$ and 
$750<i<1000$. Parameters: $\sigma_1=0.39$, $\sigma_2=0.1$, $P=320$, $R=190$, and $\gamma=0.110$.}
\label{solitary_state_chimera-2}
\end{figure}

The novel type of chimera state (Fig.~\ref{solitary_state_chimera}(a)) has been revealed  for the 
inertial coupling \eqref{inerz_eq} in the network \eqref{main_eq} and is characterized by high 
sensitivity 
to the choice of the system parameters and initial conditions. However, this chimera structure can also 
be observed for the 
dissipative coupling \eqref{dissipative_eq} and appears to be less sensitive (more robust) with respect 
to the variation of the system \eqref{main_eq} parameters. The regime of solitary state chimera realized 
for the dissipative coupling \eqref{dissipative_eq} and  the corresponding  cross-correlation 
coefficient $R_{1,i}$ for the Henon map ensemble are exemplified in 
Fig.~\ref{solitary_state_chimera-2}.

\subsection{Mutual synchronization of spatiotemporal structures}

As our numerical studies have shown, the coupled ensembles \eqref{main_eq} can demonstrate another 
interesting and important phenomenon which is very typical in coupled systems. Varying the coupling 
strength $\gamma$ between the rings often leads to the identity of spatiotemporal structures which are 
observed in the ensemble of Henon maps and the ensemble of Lozi maps. We  mean here a possible mutual 
synchronization of the spatiotemporal dynamics of the structures in these networks. Let us examine this 
issue in more detail.

\begin{figure}
\begin{tabular}{>{\centering\arraybackslash}m{1in} >{\centering\arraybackslash}m{1in} >{\centering\arraybackslash}m{1in} >{\centering\arraybackslash}m{0in}}
\includegraphics[width=.31\columnwidth]{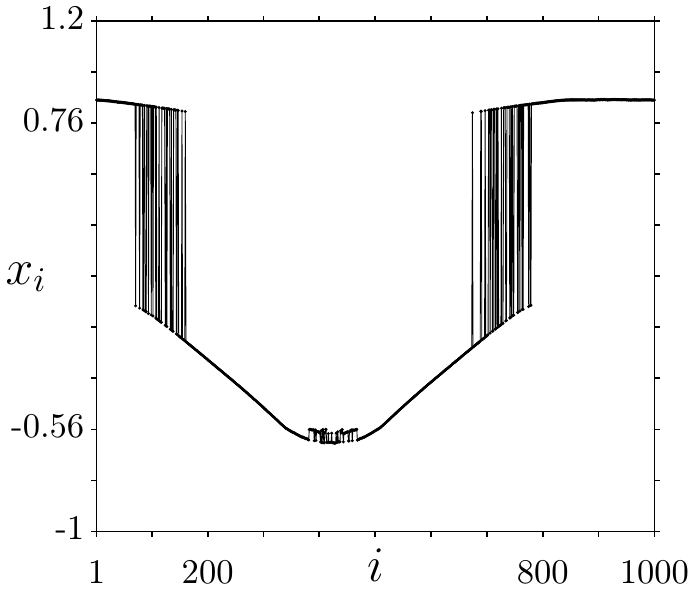} &
\includegraphics[width=.31\columnwidth]{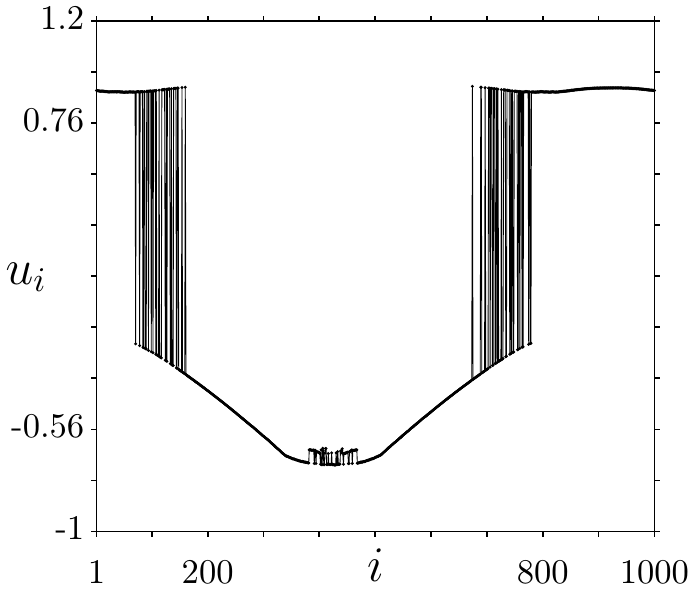}  &
\includegraphics[width=.31\columnwidth]{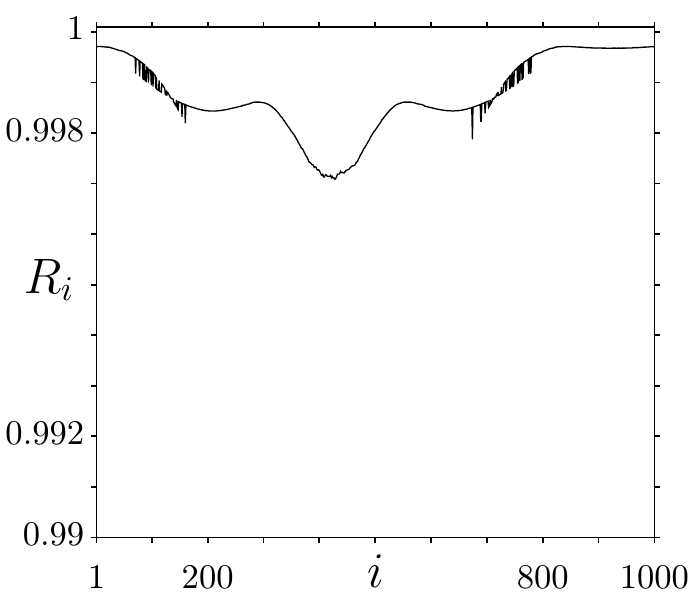} & \footnotesize (a) \\
\includegraphics[width=.31\columnwidth]{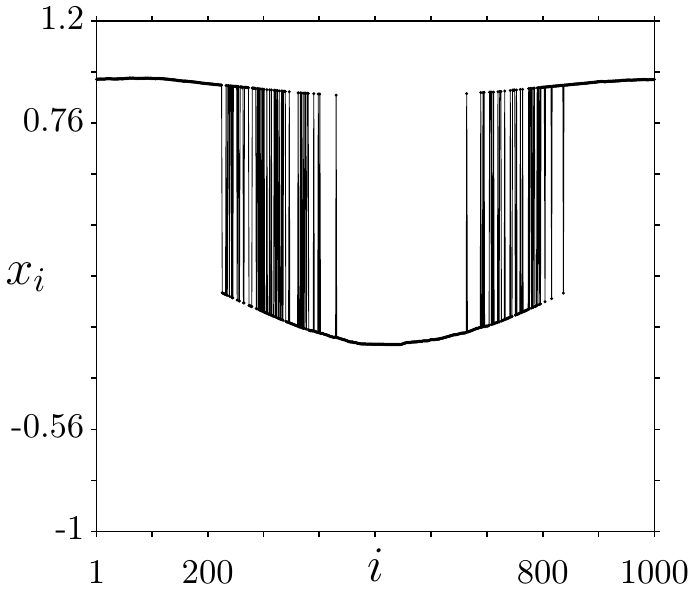} &
\includegraphics[width=.31\columnwidth]{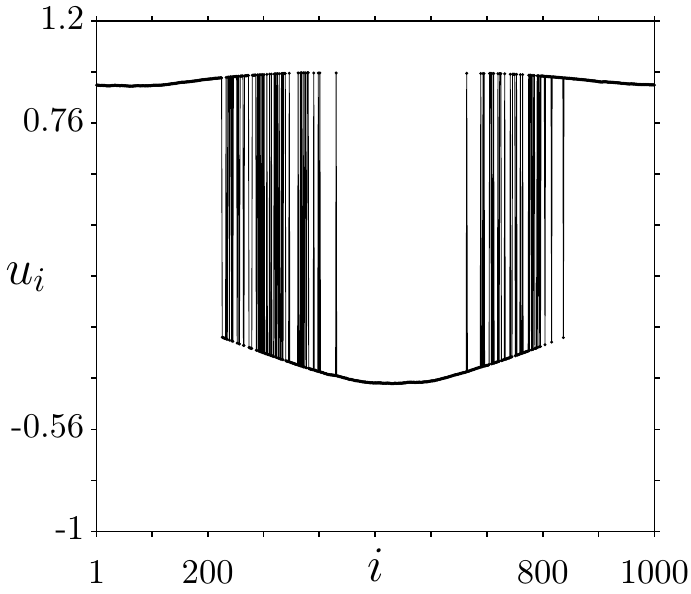}  &
\includegraphics[width=.31\columnwidth]{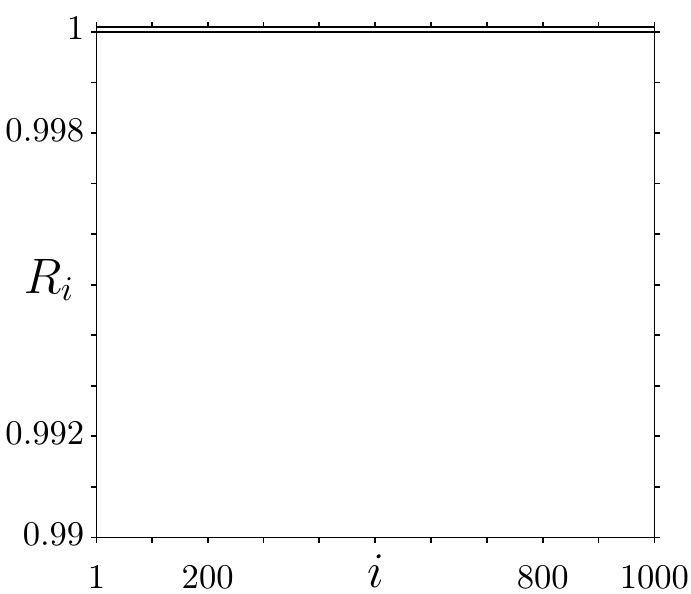} & \footnotesize (b) \\
\includegraphics[width=.31\columnwidth]{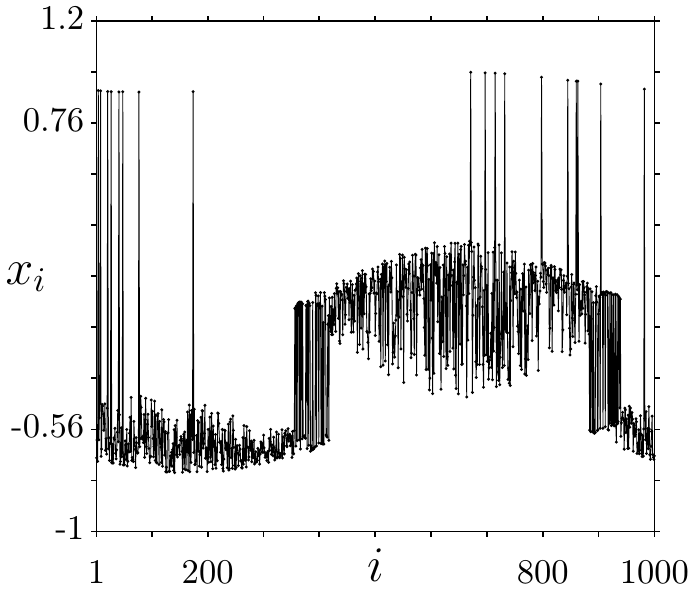} &
\includegraphics[width=.31\columnwidth]{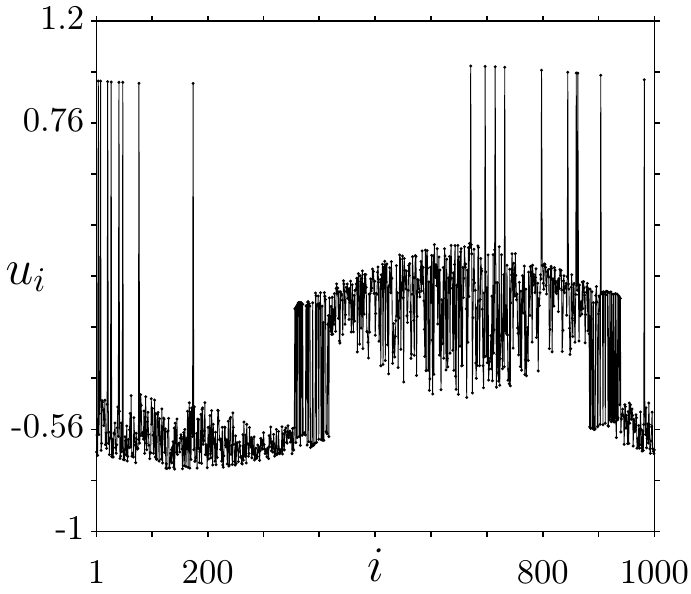}  &
\includegraphics[width=.31\columnwidth]{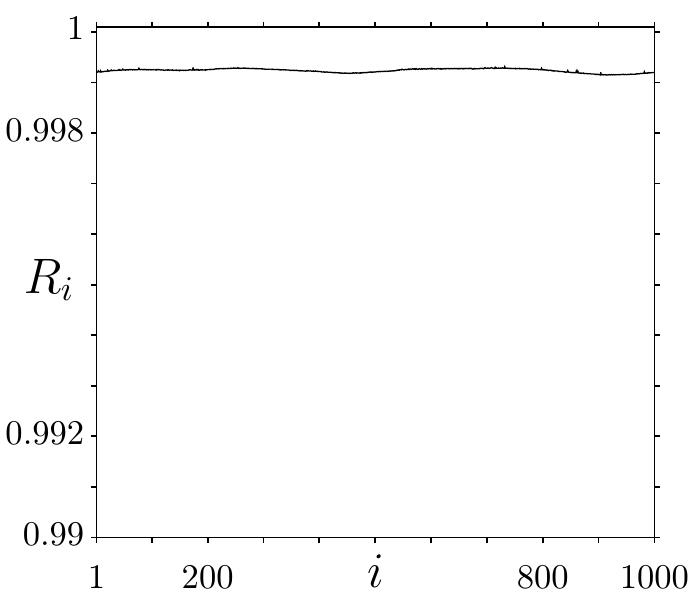} & \footnotesize (c)
\end{tabular}
\caption{Synchronous spatiotemporal structures in the network \eqref{main_eq} for the dissipative 
coupling \eqref{dissipative_eq}: (a) amplitude and phase chimeras, and (b) phase chimera at 
$\sigma_1=0.32$, $\gamma=0.375$, and (c) complex spatiotemporal structure at 
$\sigma_1=0.18$, $\gamma=0.375$. Snapshots of the dynamics of the coupled  Henon maps and of the
coupled Lozi maps are displayed in the left and middle columns, respectively, and the mutual correlation 
coefficient $R_i$ is pictured in the right column. Other parameters: $P=320$, $R=190$, 
$\sigma_2=0.15$.}
\label{sync_dif_cases}
\end{figure}

Figure~\ref{sync_dif_cases} illustrates typical examples of different identical structures in 
the system of coupled ensembles \eqref{main_eq} for the dissipative coupling 
\eqref{dissipative_eq}. Snapshots for the dynamics of the ensemble of Henon maps and of the ensemble 
of Lozi maps are shown in  the left and middle columns, respectively. In order to argue that  indeed 
we deal with  the effect of mutual
synchronization of spatiotemporal structures, we need  to justify that the following two conditions 
are fulfilled \cite{Pikovsky2001Synchro,Balanov2009Synchro}: (i) the identity of oscillatory processes in 
the corresponding oscillators of 
synchronous structures is quantified, and (ii)  the synchronization effect is realized in a finite 
region of the control parameters of the network \eqref{main_eq}. 

The identity of the oscillatory processes in the coupled ensembles of Henon and Lozi maps is diagnosed 
by calculating the cross-correlation coefficient $R_i$:
\begin{equation}
\label{inter_correlation}
\begin{aligned}
& R_i=\dfrac
{\langle \tilde x_i (t) \tilde u_i (t) \rangle}
{\sqrt{\langle \tilde x_i^2 (t) \rangle \cdot
\langle \tilde u_i^2  (t) \rangle}},
\\ & \tilde x_i = x_i - \langle x_i \rangle; \; \tilde u_i = u_i - \langle u_i \rangle.
\end{aligned}
\end{equation}
Unlike  $R_{1,i}$, the value of $R_i$ chcracterizes  the mutual correlation between the $i$th elements 
of the ensembles of Henon and Lozi maps. 
The calculation results for $R_i$ for the exemplary structures are presented in 
Fig.~\ref{sync_dif_cases} in the right column. As can be seen from the 
plots, the values of  $R_i$ are very close to unity, i.e., $0.99 \leqslant R_i \leqslant 1.0$, in all 
the three cases. This finding indicates that the oscillations of the elements in the coupled Henon and 
Lozi maps are practically identical for the spatiotemporal structures shown in 
Fig.~\ref{sync_dif_cases}. We note that the first two modes (Fig.~\ref{sync_dif_cases}(a,b)) 
correspond to the effect of mutual synchronization of chimera structures, while the latter 
(Fig.~\ref{sync_dif_cases}(c)) illustrates the mutual synchronization of a more complex spatiotemporal 
structure. As is known, the 
equality to unity characterizes the effect of complete synchronization, which is possible only in the 
case when interacting oscillators are identical.  In our case the oscillators are different 
(non-identical) and the fact that $R_i$ is close to unity testifies to nearly identical processes in the 
relevant oscillators.

\begin{figure}
\begin{center}
\includegraphics[width=1.\columnwidth]{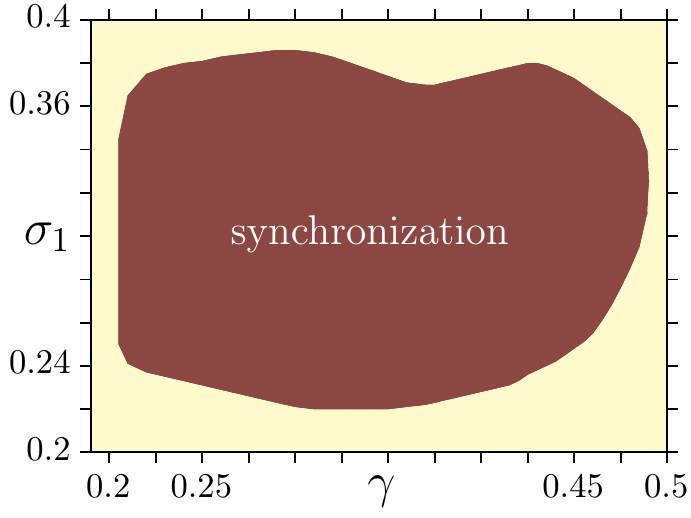}
\end{center}
\caption{Synchronization region in the $(\gamma,\sigma_1)$ parameter plane for the spatiotemporal 
structure shown in Fig.~\ref{sync_dif_cases}(a). Parameters:  $\sigma_2=0.150$, $P=320$, $R=190$.}
\label{sync_area}
\end{figure}

In order to justify that the synchronization effect is observed in a finite range of parameters' 
values of the network \eqref{main_eq}, we construct the synchronization region in the parameter plane of 
the system under study. Figure~\ref{sync_area} shows the region of synchronization in  the 
$(\gamma,\sigma_1)$ parameter plane for the complex spatiotemporal structure shown in 
Fig.~\ref{sync_dif_cases}(a). As can be seen, the phenomenon of mutual synchronization is implemented in 
a rather wide range of changes in the coupling parameters. Similar results have also been 
obtained for the region of synchronization in the plane of the other coupling parameters. 

\section{Conclusions}
\label{conclusions}

We have studied numerically the spatiotemporal dynamics of the network made of two coupled 
one-dimensional ensembles of nonlocally coupled chaotic Henon and 
Lozi maps. Our simulations have shown that for both dissipative and inertial couplings between the 
elements in the coupled ensembles \eqref{main_eq} ($\gamma>0$), all the spatiotemporal structures that
occur in  the individual  ensembles  can be realized, e.g., 
phase and amplitude chimeras, traveling waves and solitary states. The amplitude chimera regime in the 
network  \eqref{main_eq}, as well as in the individual ensemble of coupled Henon maps, is nonstationary 
and demonstrates irregular temporal switchings between amplitude and phase chimeras. The lifetime of 
the amplitude chimera in \eqref{main_eq} is, as a rule, finite and, as our studies have demonstrated, it 
can be controlled by varying the coupling coefficient $\gamma>0$  in  fairly wide ranges. 

A novel type of chimera state called  a solitary state chimera  has been revealed in the network
\eqref{main_eq} for both  the dissipative and inertial couplings between the ensembles of coupled 
Henon 
and Lozi maps. This chimera state is characterized by the coexistence of an incoherence cluster with 
uncorrelated chaotic oscillations of the cluster elements and a coherence cluster with synchronous 
chaotic oscillations. 

In conclusion, we have established the effect of mutual synchronization of various spatiotemporal 
structures in the coupled ensembles \eqref{main_eq}. The synchrony of oscillations in the 
elements  of the considered structures has been verified by calculating the cross-correlation 
coefficient $R_i$ 
\eqref{inter_correlation}, whose values in the synchronization mode are  close to unity. 
Additionally, we have constructed the region of synchronization for a selected spatiotemporal 
structure in the plane of coupling 
parameters of the  network under study and this result also corroborates the fact that 
the effect of 
synchronization is realized in the multilayer system  \eqref{main_eq}.

\section{Acknowledgments}
We acknowledge support by DFG in the framework of SFB 910 and the Russian Science Foundation (grant No. 
16-12-10175).

\bibliographystyle{apsrev}

\end{document}